\documentclass[aps,graphicx,10pt,showkeys,showpacs,twocolumn]{revtex4}
\usepackage{amsmath}
\usepackage{amscd}
\usepackage{graphicx}
\usepackage{subfigure}
\usepackage{appendix}
\usepackage{multirow}
\usepackage{mathrsfs}
\usepackage{CJK}

\begin{document}

\begin{CJK*}{GBK}{song}
\CJKtilde\CJKindent

\title[Short Title]{Parameter trajectory engineering for state transfer and quantum sensing in non-Hermitian two-level systems}

\author{Qi-Cheng Wu$^{1,2,3,}$\footnote{E-mail: wuqi.cheng@163.com}}
\author{Yan-Hui Zhou$^{1,3}$}
\author{Biao-liang Ye$^{1}$}
\author{Tong Liu$^{1}$}
\author{Yi-Hao Kang$^{2}$}
\author{Qi-Ping Su$^{2}$}
\author{Chui-Ping Yang$^{2,}$\footnote{E-mail: yangcp@hznu.edu.cn}}

\affiliation{$^{1}$Quantum Information Research Center and Jiangxi Province Key Laboratory of Applied Optical Technology, Shangrao Normal University, Shangrao 334001, China\\
$^{2}$School of Physics, Hangzhou Normal University, Hangzhou
311121, China\\
$^{3}$Institute of Quantum Science and Technology, Yanbian
University, Yanji, Jilin 133002, China}

\begin{abstract}
Exceptional points (EPs) in non-Hermitian systems give rise to
enhanced sensitivity and chiral state transfer, which are
important for quantum technologies. Although parameter
trajectories encircling EPs can control symmetric and chiral state
transfer, their robustness against practical perturbations and
their role in quantum sensing remain largely unexplored. Here, we
study three time-modulated parameter loops in a non-Hermitian
two-level system to show how trajectory design governs
state-transfer symmetry, robustness, and sensing performance.
Trajectories avoiding the EP support robust symmetric transfer,
while those encircling the EP yield chiral transfer governed by
the topological winding number $\nu=\pm1/2$, whose robustness
depends on the distance to the EP and the encircling direction.
For quantum sensing, trajectory engineering enables tuning of
sensitivity amplitude, time window, and parameter selectivity in
both eigenvalue-based and eigenstate-based sensors. Notably,
eigenstate-based sensing achieves full parameter selectivity that
is unattainable with eigenvalue-based methods. Our results
establish a quantitative connection between trajectory topology
and system dynamics, providing a unified framework for robust
state-transfer protocols and high-performance quantum sensors.
\end{abstract}

\pacs {03.67.-a, 42.50.Dv, 74.40.Kb} \keywords{Exceptional points;
Non-Hermitian systems; Quantum sensing; State transfer}

\maketitle

\section{Introduction}

Since the discovery of exceptional points (EPs) as spectral
singularities in non-Hermitian (NH) systems, research in NH
quantum dynamics has revealed many counterintuitive phenomena. At
an EP, both eigenvalues and their corresponding eigenvectors
coalesce. These phenomena have far-reaching implications for
quantum
technologies~\cite{EPs1,EPs2,EPs3,square-root1,square-root2}. They
include enhanced sensitivity to external
perturbations~\cite{Enhanced-sensitivity,sensing-microcavity,Microcavity-sensor}
and nontrivial topological
structures~\cite{topology1,topology-encircling}. These unique
properties stem from the spectral and eigenstate characteristics
intrinsic to NH systems~\cite{NH1,NH2,NH3}. They lay the
foundation for innovative quantum device designs.

Among the promising research directions enabled by these
properties, controlling state transfer via parameter trajectory
engineering has become a central topic. Recent studies show that
two distinct types of state transfer can be realized. These are
achieved by evolving system parameters along suitably designed
closed paths. The first type is symmetric (adiabatic) state
transfer, which is independent of the encircling
direction~\cite{topology-encircling,wu-encircling,Arkhipov-encircling2}.
The second type is asymmetric (chiral) state transfer, which is
determined solely by this direction~\cite{PhysRevLett.129.127401,
PhysRevLett.124.153903,bell-encircling,Xu-encircling,
Li-encircling,Feilhauer-encircling,Ergoktas-encircling,Arkhipov-encircling1,
Tang-encircling,Hassan-encircling2,negativecavity2}. Importantly,
whether the parameter trajectory encircles an EP directly governs
these two transfer modes~\cite{PhysRevX.8.021066}. The ability to
tailor the symmetry of state transfer through trajectory design
opens new avenues for quantum state manipulation. This has
promising applications in quantum switches, quantum memories, and
quantum sensors. However, a critical knowledge gap remains in
understanding the practical utility of these trajectories. In
realistic experimental settings, system imperfections,
environmental noise, and parameter fluctuations are unavoidable.
Yet, the robustness of symmetric and chiral state transfer against
such perturbations has not been systematically analyzed.

This gap is further compounded by the underutilization of
parameter trajectory engineering in quantum sensing. EPs are well
known to boost quantum sensing sensitivity by inducing divergent
susceptibilities~\cite{Arkhipov2026,quantum-sensing2,quantum-sensing3,quantum-sensing4,quantum-sensing5,quantum-sensing6,sensing-add1,experiment-realize3}.
However, the role of trajectory design in tuning key sensor
performance metrics remains largely unexplored. These metrics
include sensitivity amplitude, operating time windows, and
parameter selectivity. This oversight limits the practical
deployment of EP-enhanced quantum sensors. Optimal performance
often requires precise control over these metrics.

The need to address these gaps is further underscored by the
topological complexity of EPs in two-level NH models. Encircling a
single EP typically induces a square-root branch cut in the
eigenenergies. This is accompanied by eigenstate permutation and
the accumulation of a \(\pi\) Berry
phase~\cite{square-root1,square-root2,PhysRevLett.127.253901}.
These intertwined effects reveal that the topological
characterization of EPs cannot rely solely on eigenvalue spectra.
The internal structure of eigenstates and their associated
geometric phases must also be considered. This complexity
highlights the importance of understanding how trajectory design
governs both state-transfer dynamics and sensing performance. The
trajectory topology directly shapes the topological properties
that dictate the overall system behavior.

To fill these critical gaps, we systematically investigate the
regulatory effects of parameter trajectories. We focus on
state-transfer symmetry, robustness, and quantum-sensing
performance in a NH two-level system. We define three
representative time-modulated parameter loops. They differ in EP
encirclement and the modulation of key system parameters, such as
detuning, coupling strength, and loss. By analyzing the
state-transfer fidelity, topological winding number, and
susceptibility of sensing observables, we establish a clear
quantitative link between trajectory topology and system response.
Specifically, this work aims to achieve three main goals. (1) To
clarify how trajectory design controls the transition between
symmetric and chiral state transfer, as well as the robustness of
each transfer mode against external perturbations. (2) To
demonstrate how trajectory engineering can tailor the performance
of both eigenvalue-based (spectral-shift detection) and
eigenstate-based (population-monitoring) quantum sensors. (3) To
derive practical design principles for optimizing NH
state-transfer protocols and quantum sensors. These findings not
only advance the fundamental understanding of NH system dynamics
but also provide actionable guidance for developing robust,
high-performance quantum technologies.

The remainder of this paper is organized as follows:
Sec.~\ref{sec:model} introduces the NH two-level Hamiltonian and
the three representative parameter trajectories.
Sec.~\ref{sec:stateTransfer} analyzes the regulatory effect of
trajectories on state transfer symmetry and robustness.
Sec.~\ref{sec:sensing} explores the impact of trajectory design on
eigenvalue- and eigenstate-based quantum sensing. Finally,
Sec.~\ref{sec:conclusion} summarizes our key findings and
discusses future perspectives.

\section{Non-Hermitian two-level Hamiltonian and parameter trajectory}
\label{sec:model}

\subsection{Non-Hermitian two-level Hamiltonian}

Our analysis is based on a non-Hermitian two-level Hamiltonian
(setting $\hbar=1$)~\cite{JC2,wu-encircling1},
\begin{equation}\label{eq:Heff}
H=\begin{pmatrix}
\omega_{a}+\delta-\dfrac{i\gamma}{2} & g \\[6pt]
g & \omega_{a}+\dfrac{i\gamma}{2}
\end{pmatrix},
\end{equation}
where $\omega_{a}$ corresponds to the ground-state energy,
$\delta$ represents the energy detuning between the two levels,
$\gamma$ is the dissipation coefficient of the system, and $g$
denotes the coupling strength between the two energy levels.

The eigenvalues of $H$ are
\begin{subequations}\label{eq:eigenvalues}
\begin{align}
E_{\pm}&=\frac{1}{2}\bigl[2\omega_{a}+\delta\pm\Delta_E\bigr], \\
\Delta_E&=\frac{i}{2}\sqrt{(2\gamma+2i\delta)^2-16g^{2}},
\end{align}
\end{subequations}
and the corresponding right eigenvectors are
\begin{equation}\label{eq:rightEvecs}
|\phi_{\pm}\rangle=\frac{1}{S_{\pm}}\bigl[A_{\pm},\,4g\bigr]^{\mathsf
T},
\end{equation}
with $A_{\pm}=2\delta-2i\gamma\pm2\Delta_E$ and
$S_{\pm}=\sqrt{|A_{\pm}|^2+|4g|^2}$. The left eigenvectors are
\begin{equation}\label{eq:leftEvecs}
\langle\tilde{\phi}_{\pm}|=\frac{1}{S_{\pm}}\bigl[A^{*}_{\pm},\,4g\bigr].
\end{equation}
The biorthonormal partners satisfy
$\langle\tilde{\phi}_{n}|\phi_{m}\rangle=\delta_{nm}$ and the
completeness relations
$\sum_{n}|\tilde{\phi}_{n}\rangle\langle\phi_{n}|
=\sum_{n}|{\phi}_{n}\rangle\langle\tilde{\phi}_{n}|=1$ \cite{NH3}.

\subsection{Parameter trajectory}
\label{sec:EPtraj}

We consider the following time-modulated parameter loop
\cite{wu-encircling,bell-encircling,quantum-sensing5}:
\begin{equation}
\begin{aligned}
\delta(\theta) &= \Delta_{0}\sin\theta, \\
g(\theta)     &= g_{0}+G_{0}\cos\theta, \\
\gamma(\theta)&= \Gamma_{0}\sin^2(\theta/2),
\end{aligned}
\label{eq:traj}
\end{equation}
where $\Delta_{0},g_{0},G_{0},\Gamma_{0}\in {R}$ are constants and
$\theta=\omega t$ denotes the angular parameter. The sinusoidal
modulation is adopted for analytical tractability and clear
demonstration of dynamical effects. The qualitative behavior,
particularly the emergence of optimal temporal windows for
enhanced sensitivity, is expected to hold for other smooth
modulation schemes.

We focus on three representative trajectories: Trajectory 1
[$\delta\equiv0$, modulated $g(\theta)$ and $\gamma(\theta)$],
\begin{equation}
\begin{aligned}
\delta(\theta)&\equiv0, \\
g(\theta)&=0.01+0.2\cos\theta, \\
\gamma(\theta)&=0.2\sin^2(\theta/2),
\end{aligned}
\label{eq:traj1}
\end{equation}
where $\Delta_0=0$, $g_0=0.01$, $G_0=\Gamma_0=0.2$. Trajectory 2
[constant $g$, modulated $\delta(\theta)$ and $\gamma(\theta)$],
\begin{equation}\label{eq:traj2}
\begin{aligned}
\delta(\theta)&=0.04\sin\theta,\\
 g(\theta)&\equiv0.1,\\
\gamma(\theta)&=0.1\sin^2(\theta/2),
\end{aligned}
\end{equation}
where $\Delta_0=0.04$, $g_0=0.1$, $G_0=0$, $\Gamma_0=0.1$.
Trajectory 3 [constant $\gamma$, modulated $\delta(\theta)$ and
$g(\theta)$],
\begin{equation}\label{eq:traj3}
\begin{aligned}
\delta(\theta)&=0.2\sin\theta,\\
 g(\theta)&=0.2+0.2\cos\theta,\\
\gamma(\theta)&\equiv0.1,
\end{aligned}
\end{equation}
where $\Delta_0=g_0=G_0=0.2$.

\section{Regulatory effect of parameter trajectory on state transfer}\label{sec:stateTransfer}

Recent studies have shown that both the symmetric (adiabatic)
state transfer
\cite{topology-encircling,Arkhipov-encircling2,wu-encircling} and
asymmetric (chiral) state transfer
\cite{PhysRevLett.129.127401,Feilhauer-encircling} can be realized
by evolving parameters along loops that encircle an EP. In this
section, we systematically investigate how the choice of parameter
trajectory influences the symmetry and robustness of state
transfer.

\subsection{Symmetric and asymmetric transfer controlled by trajectory}\label{subsec:transferSym}

\begin{figure}
\centering \scalebox{0.43}{\includegraphics{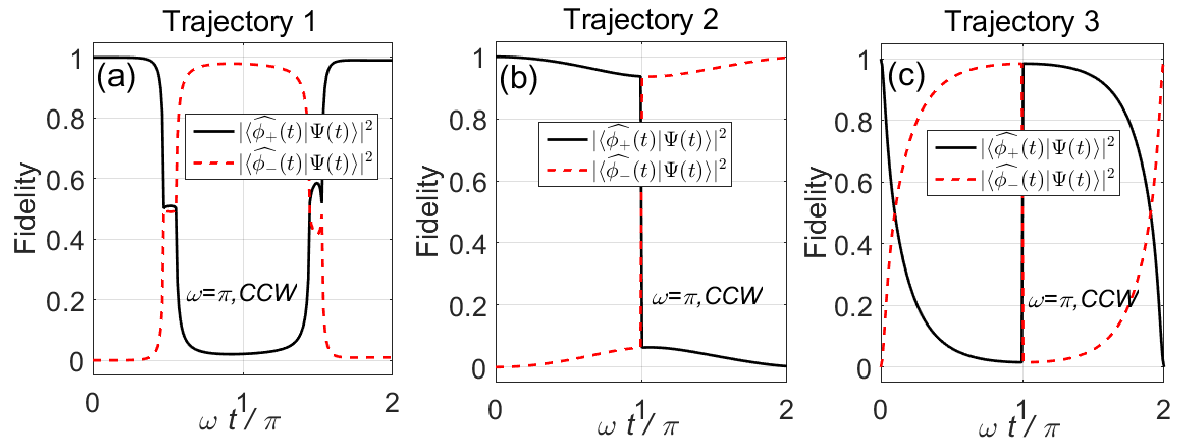}}
\scalebox{0.43}{\includegraphics{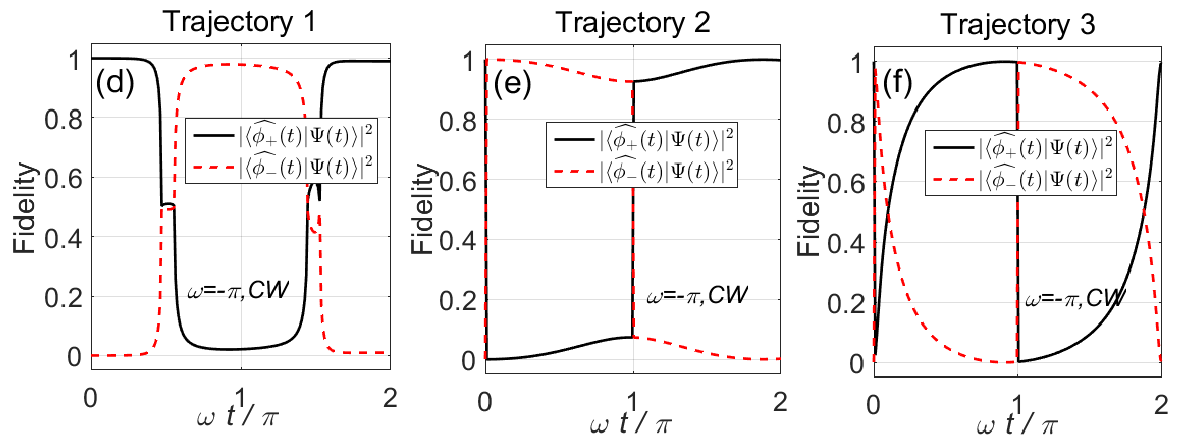}} \caption{Time
evolution of the fidelities $F_{\pm}(t)$ for the three
trajectories under (a)-(c) CCW and (d)-(f) CW encirclement. The
initial state is $|\phi_{+}(0)\rangle$, and the trajectories'
parameters have been set in
Eqs.~(\ref{eq:traj1})-(\ref{eq:traj3}).} \label{fig:fidelity}
\end{figure}
To quantify the state-transfer performance, we employ the fidelity
with respect to the instantaneous right eigenstates
$|\phi_{m}(t)\rangle$ ($m=+,-$):
\begin{equation}\label{eq:fidelity}
F_{m}(t)=\bigl|\langle\tilde{\phi}_{m}(t)|\Psi(t)\rangle\bigr|^{2},
\end{equation}
where $|\Psi(t)\rangle$ is the time-evolved state obtained by
numerically integrating the Schr\"odinger equation
\begin{equation}\label{eq:schrodinger}
i\partial_{t}|\Psi(t)\rangle=H(t)|\Psi(t)\rangle,
\end{equation}
and the system is initialized in
$|\Psi(0)\rangle=|\phi_{+}(0)\rangle$.

Figure~\ref{fig:fidelity} shows $F_{\pm}(t)$ for the three
trajectories traversed in both counterclockwise (CCW,
$\theta:0\to2\pi$) and clockwise (CW, $\theta:0\to-2\pi$)
directions. For Trajectory~1 [Figs.~\ref{fig:fidelity}(a) and
(d)], the fidelities exhibit symmetric behavior. After a half
period $T/2=\pi/\omega$, the states $|\phi_{+}\rangle$ and
$|\phi_{-}\rangle$ are exchanged. They return to the initial
configuration after a full period $T=2\pi/\omega$, irrespective of
the encircling direction. This direction independence confirms the
adiabatic and symmetric nature of the transfer. In contrast,
Trajectories~2 and~3 [Figs.~\ref{fig:fidelity}(b),(e) and (c),(f)]
display pronounced chirality. For CCW encirclement, the
eigenstates are exchanged after a full cycle. For CW encirclement,
they return to their initial forms. This winding-direction
dependence enables asymmetric (chiral) state
switching~\cite{PhysRevLett.129.273601,PhysRevLett.132.243802}.
These results demonstrate that by appropriately designing the
parameter loop, one can flexibly achieve either symmetric or
asymmetric transfer. This is a key capability for quantum state
manipulation in NH systems.

\subsection{Robustness of transfer against parameter variations}
\label{subsec:robustness}

\begin{figure}
\centering \scalebox{0.45}{\includegraphics{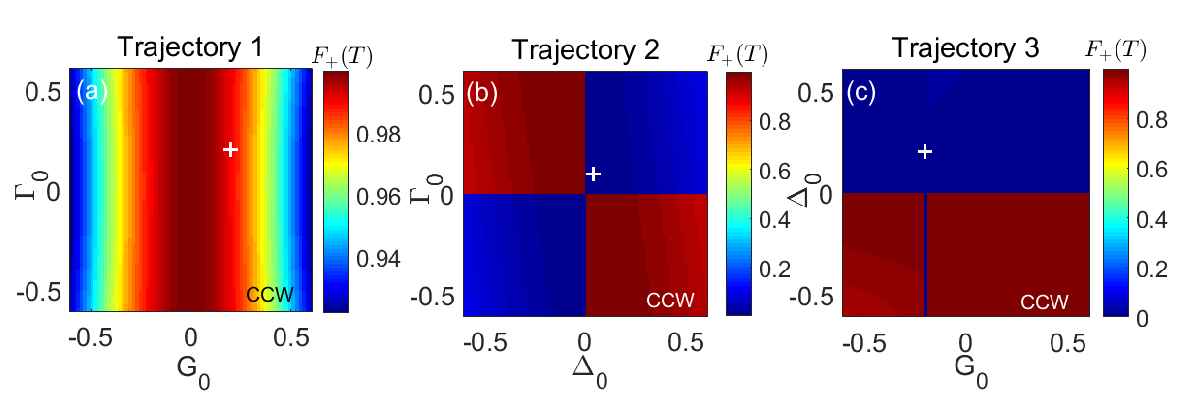}}
\scalebox{0.45}{\includegraphics{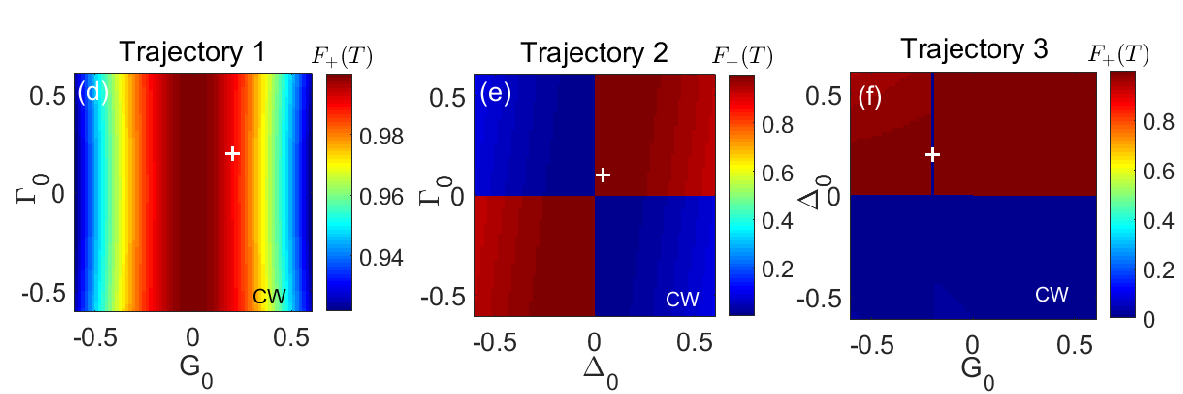}}
\caption{Final fidelity $F_{+}(T)$ after a full period
$T=2\pi/\omega$ for the three trajectories under (a)-(c) CCW and
(d)-(f) CW encirclement. The white cross marks the starting/ending
point and the initial state is $|\phi_{+}(0)\rangle$. }
\label{fig:fidelity_var}
\end{figure}

\begin{figure}
\centering \scalebox{0.45}{\includegraphics{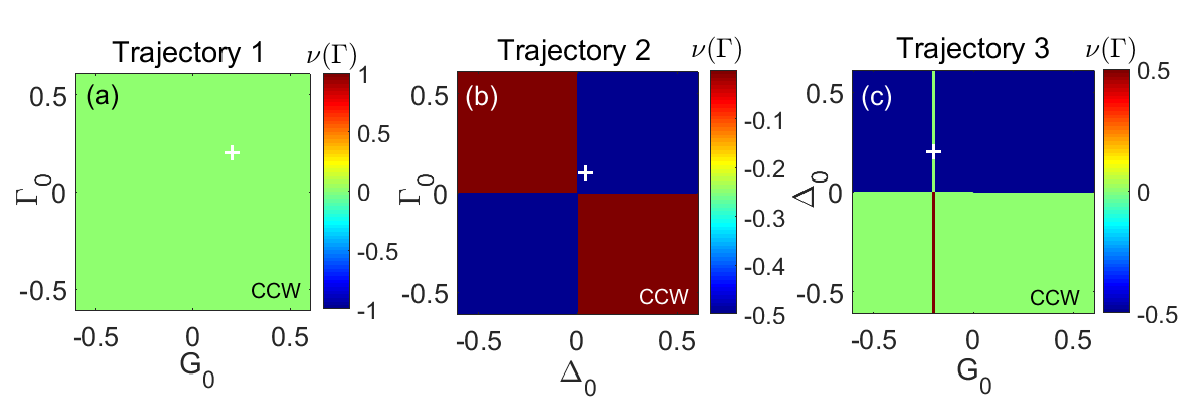}}
\scalebox{0.45}{\includegraphics{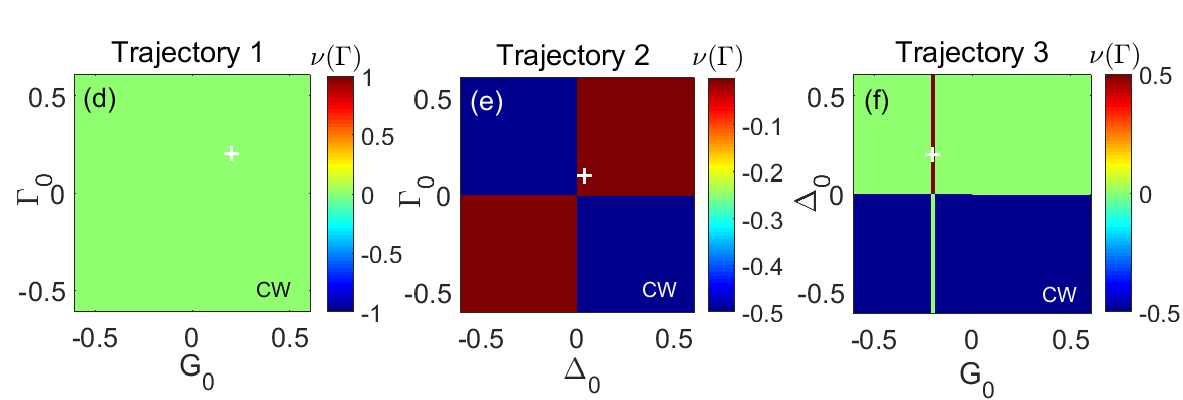}} \caption{Winding
number $\nu(\Gamma)$ for the three trajectories under (a)-(c) CCW
and (d)-(f) CW encirclement. The white cross marks the same
starting point as in Fig.~\ref{fig:fidelity_var}.} \label{fig:TP}
\end{figure}

Practical implementations are inevitably subject to external
perturbations. It is therefore crucial to understand how
variations of the trajectory parameters affect the fidelity. We
also need to determine whether the observed chiral or non-chiral
behavior remains stable. A useful tool is the vorticity (winding
number) associated with a closed loop $\Gamma$ in parameter space
\cite{tp1,tp2}:
\begin{equation}\label{eq:vorticity}
\nu(\Gamma)=-\frac{1}{2\pi}\oint_{\Gamma}
\nabla_{\theta}\arg\bigl[E_{+}(\theta)-E_{-}(\theta)\bigr]\cdot
d\theta .
\end{equation}
Reparametrizing the complex energy splitting as $E_{+}-E_{-}=
\sqrt{r}\,e^{i\theta/2}$ yields $\nu(\Gamma)=\pm1/2$ for a loop
that encircles an EP once \cite{square-root1,square-root2}. This
topological invariant directly links the loop geometry to the
dynamical outcome. We numerically calculate the final fidelity
\(F_{+}(T)\) following a full period \(T=2\pi/\omega\) for the
three trajectories. We systematically vary the relevant control
parameters: \(G_0\) and \(\Gamma_0\) for Trajectory 1,
\(\Delta_0\) and \(\Gamma_0\) for Trajectory 2, and \(\Delta_0\)
and \(G_0\) for Trajectory 3. The results of these numerical
simulations are presented in Fig.~\ref{fig:fidelity_var}, and
corresponding maps of the winding number \(\nu(\Gamma)\) are
displayed in Fig.~\ref{fig:TP}.

For Trajectory 1 [Figs.~\ref{fig:fidelity_var}(a),(d) and
\ref{fig:TP}(a),(d)], the final fidelity \(F_{+}(T)\) remains
close to unity (\(\approx0.99\)) across the entire red range. It
shows no dependence on the direction of parameter encirclement.
This result demonstrates that the system returns to its initial
eigenstate \(|\phi_+(T)\rangle\) after one complete cycle of
parameter evolution. It exhibits symmetric (non-chiral) state
transfer, where the outcome is insensitive to the direction of the
encircling path.

Consistently, the topological invariant \(\nu(\Gamma)\) equals 0
uniformly throughout the entire parameter space, irrespective of
the encircling direction. This behavior arises because the
parameter loop never encloses the EP. It renders the dynamics
topologically trivial and highly robust against parameter
fluctuations. This is an essential feature for applications
requiring stable, direction-insensitive state transfer. As will be
shown in Sec.~\ref{sec:sensing}, this same topological triviality
also underlies the predictable, smoothly tunable response of
Trajectory 1 in quantum sensing applications.

Trajectory 2 [Figs.~\ref{fig:fidelity_var}(b),(e) and
\ref{fig:TP}(b),(e)] exhibits a distinct quadrantal pattern in
both the fidelity and winding number distributions. For CCW
encirclement, \(F_{+}(T)\approx1\) in the first and third
quadrants, while it drops to \(\approx0\) in the second and fourth
quadrants. For CW encirclement, this pattern is inverted.

Importantly, the regions where \(F_{+}(T)\approx0\) coincide
exactly with those where \(|\nu(\Gamma)|=1/2\). This confirms that
chiral state transfer occurs when the parameter loop winds around
the EP \cite{PhysRevLett.127.253901,PhysRevLett.129.127401}.
Notably, the transitions between \(\nu(\Gamma)=0\) and
\(|\nu(\Gamma)|=1/2\), and correspondingly between \(F_{+}=1\) and
\(F_{+}=0\), are sharp. This indicates that small perturbations in
the control parameters can switch the winding number and thus
alter the state transfer outcome. This sensitivity originates from
the proximity of the parameter loop to the EP. Minor variations in
parameters can change whether the loop encircles the EP. This
modifies both the winding number and the resulting transfer
behavior. Starting from the white cross (denoting the initial
parameter point), the CCW loop yields a non-trivial topological
value (\(\nu(\Gamma) = -0.5\)) after one full cycle. In contrast,
the CW loop results in a trivial topological value (\(\nu(\Gamma)
= 0\)) under the chosen parameters. This observation further
highlights the asymmetric and perturbation-sensitive nature of
Trajectory 2.

For Trajectory 3 [Figs.~\ref{fig:fidelity_var}(c),(f)
and~\ref{fig:TP}(c),(f)], the final fidelity \(F_+(T)\) and
winding number \(\nu(\Gamma)\) exhibit a distinct vertical banded
structure in the \((\Delta_0,G_0)\) parameter plane with
\(\Gamma_0=0.1\) fixed. The band is centered at \(G_0\approx0.2\)
and extends uniformly along the \(\Delta_0\) direction. For
counterclockwise (CCW) encirclement, \(F_+(T)\approx 0\) and
\(\nu(\Gamma)=-0.5\) inside the band, while \(F_+(T)\approx 1\)
and \(\nu(\Gamma)=0\) outside. For clockwise (CW) encirclement,
the pattern is reversed: \(F_+(T)\approx 1\) and
\(\nu(\Gamma)=+0.5\) inside the band, and \(F_+(T)\approx 0\) and
\(\nu(\Gamma)=0\) outside.

This behavior differs fundamentally from the quadrantal pattern of
Trajectory 2. The topological response here depends purely on
\(G_0\) and is uniform in \(\Delta_0\), yielding strong robustness
against detuning fluctuations. The sign reversal of
\(\nu(\Gamma)\) and the corresponding flip in \(F_+(T)\) between
CCW and CW evolution confirm that a nonzero winding number is
necessary but not sufficient for eigenstate exchange; the final
state also depends on the geometric phase accumulated along the
closed trajectory. Moreover, the vertical stripe near
\(G_0\approx0.2\) represents a sharp topological phase boundary
separating the trivial phase \(\nu(\Gamma)=0\) and the nontrivial
phase \(\nu(\Gamma)=\pm1/2\). This transition occurs because
\(G_0\approx0.2\) is the critical coupling amplitude at which the
trajectory tangentially approaches or crosses the EP-associated
branch cut~\cite{PhysRevLett.132.243802}. In this narrow stripe,
the energy splitting \(\Delta_E\) is strongly suppressed, driving
the system into the near-EP regime where rapid rearrangements of
eigenenergies and eigenstates take place. The strictly vertical
transition boundary reveals that the topological transition is
governed solely by \(G_0\) and fully decoupled from \(\Delta_0\).
Such a clean, parameter-isolated response endows Trajectory 3 with
exceptional parameter selectivity and robustness: high sensitivity
to weak variations in \(G_0\) and strong immunity to unintended
perturbations in \(\Delta_0\), which is highly desirable for
robust single-parameter quantum sensing.

These findings establish a clear quantitative connection between
the topological invariant \(\nu(\Gamma)\) and the dynamical
outcome of state transfer. (i) When \(\nu(\Gamma)=0\), i.e., the
parameter loop does not encircle the EP, the system exhibits
symmetric (non-chiral) state transfer that is robust against
parameter variations. (ii) When \(\nu(\Gamma)=\pm1/2\), i.e., the
parameter loop encircles the EP, chiral state transfer occurs. The
robustness of the transfer depends on both the proximity of the
loop to the EP and the direction of encirclement. The EP acts as a
topological defect in the parameter space \cite{topology1}.
Encircling it imprints a non-Abelian geometric phase that dictates
the final state of the system. Importantly, as long as the winding
number \(\nu(\Gamma)\) is preserved, the symmetry of the state
transfer is protected against smooth perturbations of the control
parameters.

\section{Regulatory effect of parameter trajectory on quantum sensing}
\label{sec:sensing}

Having established that parameter trajectories govern both the
symmetry and robustness of state transfer, we now explore their
impact on quantum sensing. It is well known that the divergent
susceptibility in the vicinity of an EP can enhance the
sensitivity of both eigenvalue-based and eigenstate-based
detectors
\cite{Arkhipov2026,quantum-sensing2,quantum-sensing3,Enhanced-sensitivity,Microcavity-sensor}.
Here, we demonstrate that trajectory engineering provides an
additional degree of freedom to tailor the amplitude, time window,
and selectivity of the sensor response.

\subsection{Eigenvalue-based sensing}
\label{subsec:eigenvalueSensing}

\begin{figure}[htb]
\centering \scalebox{0.45}{\includegraphics{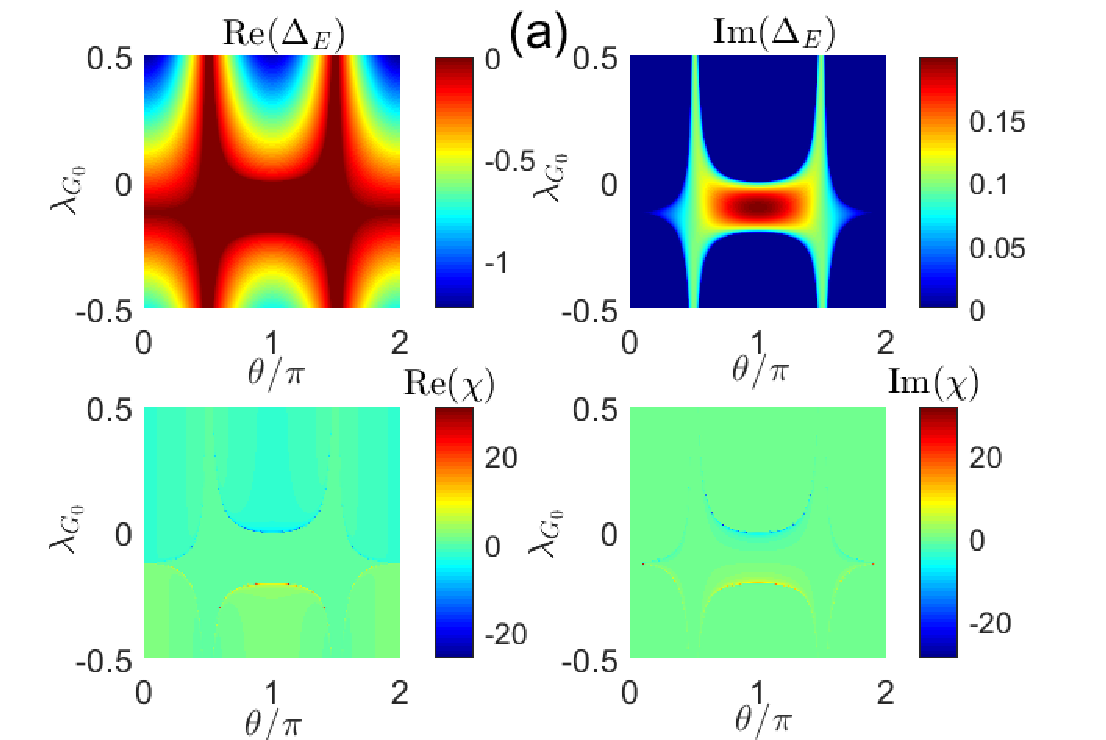}}
\scalebox{0.45}{\includegraphics{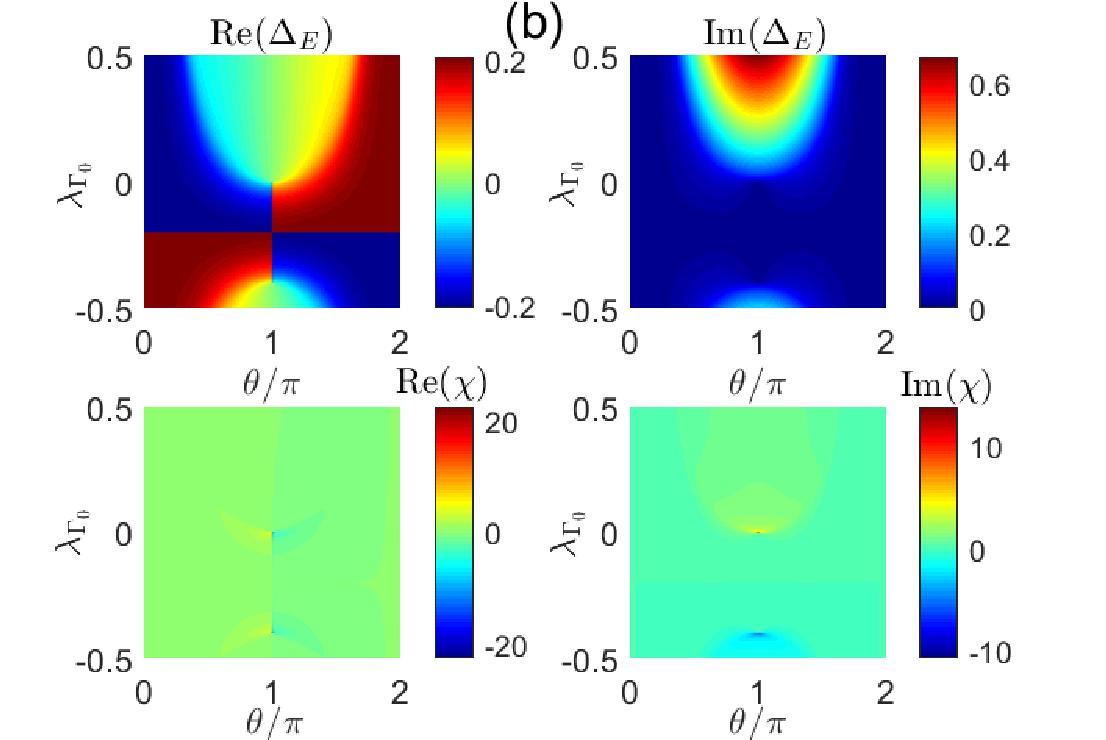}}
\scalebox{0.45}{\includegraphics{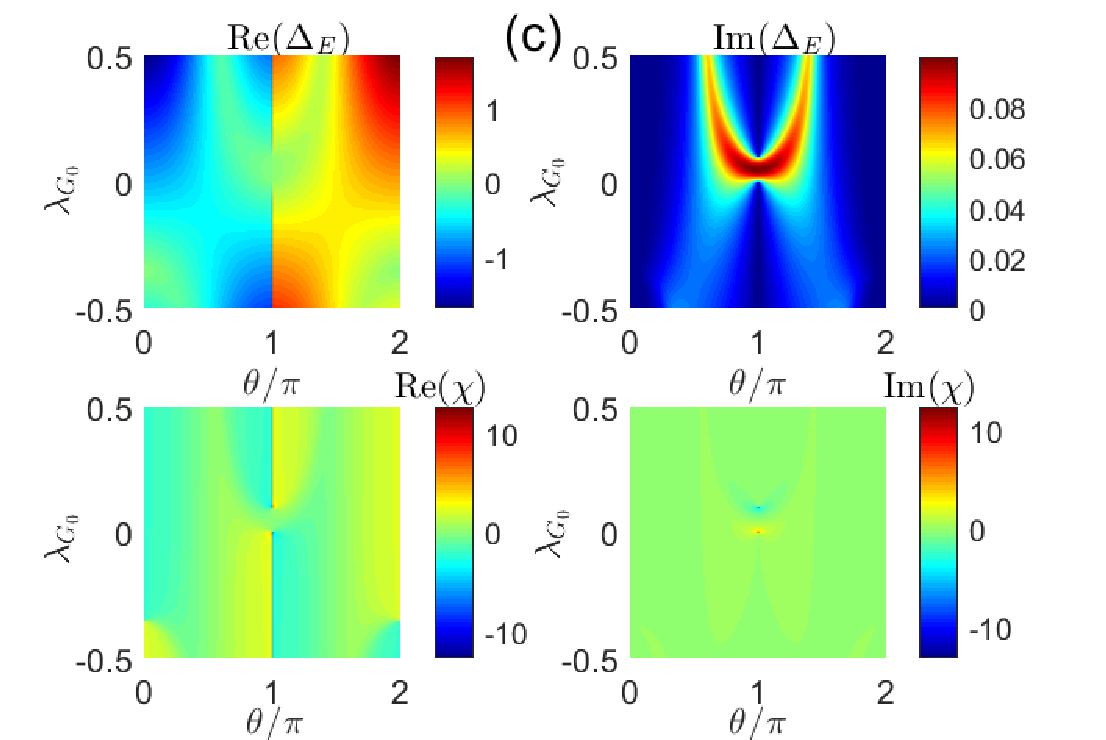}} \caption{Real
and imaginary parts of the energy splitting $\Delta_E(\theta)$ and
susceptibility $\chi(\theta)$ as functions of the scaled time
$\theta/\pi$ for small perturbations of control parameters. (a)
Trajectory 1 with $G_0$ perturbation ($G_0^{\text{I}} = 0.11$);
(b) Trajectory 2 with $\Gamma_0$ perturbation
($\Gamma_0^{\text{I}} = 0.2$); (c) Trajectory 3 with $G_0$
perturbation ($G_0^{\text{I}} = 0.15$). Other parameters are given
in Eqs.~(\ref{eq:traj1})-(\ref{eq:traj3}).} \label{fig:E}
\end{figure}

\begin{figure}[htb]
\centering \scalebox{0.45}{\includegraphics{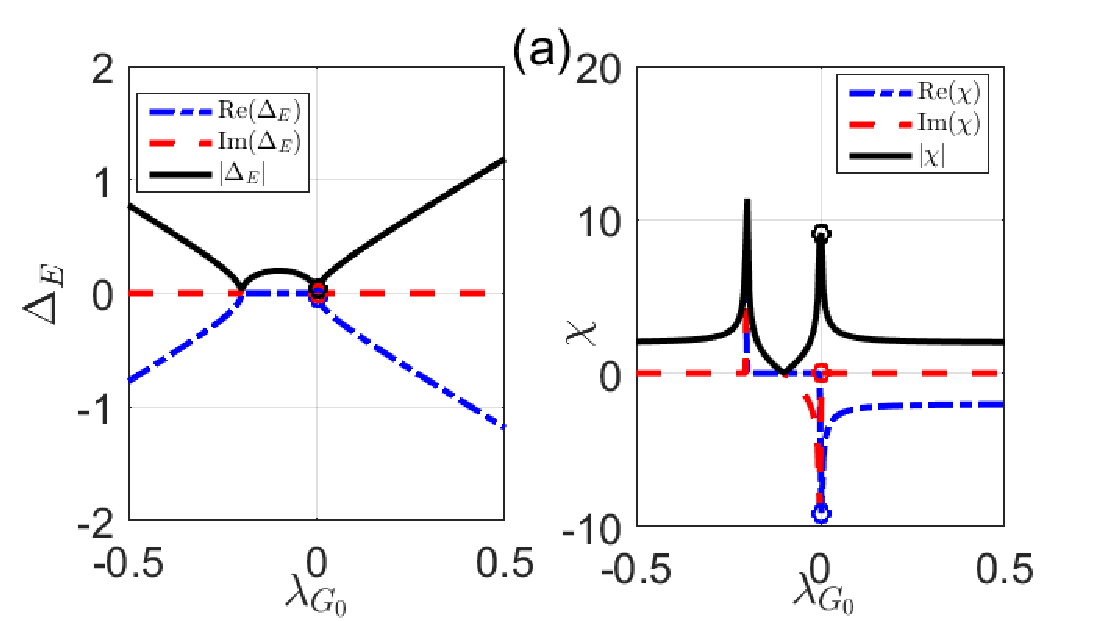}}
\scalebox{0.45}{\includegraphics{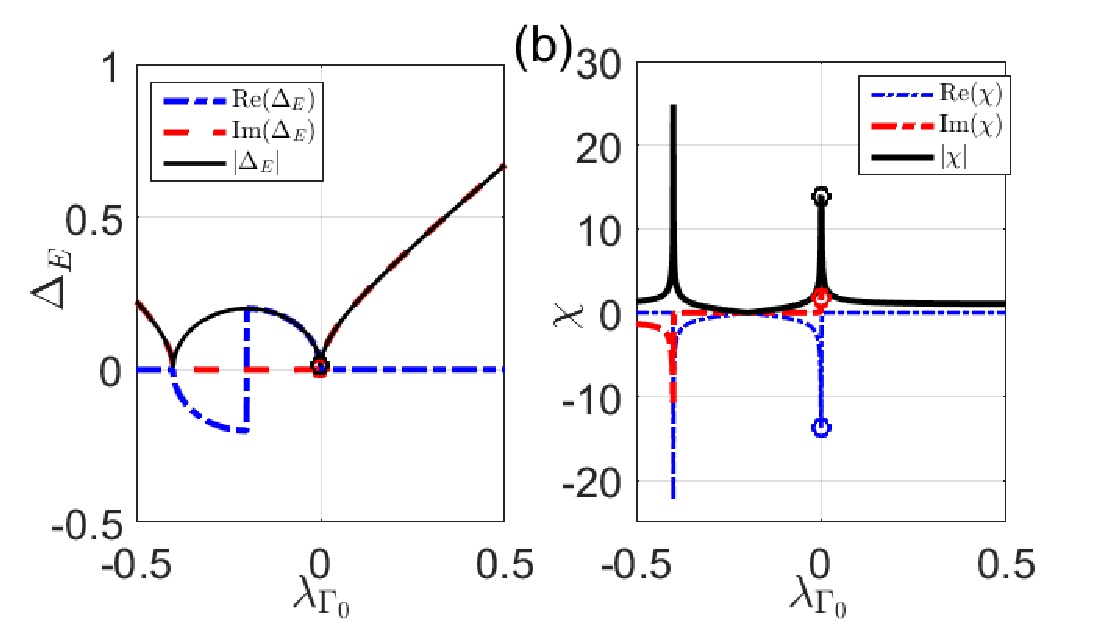}}
\scalebox{0.45}{\includegraphics{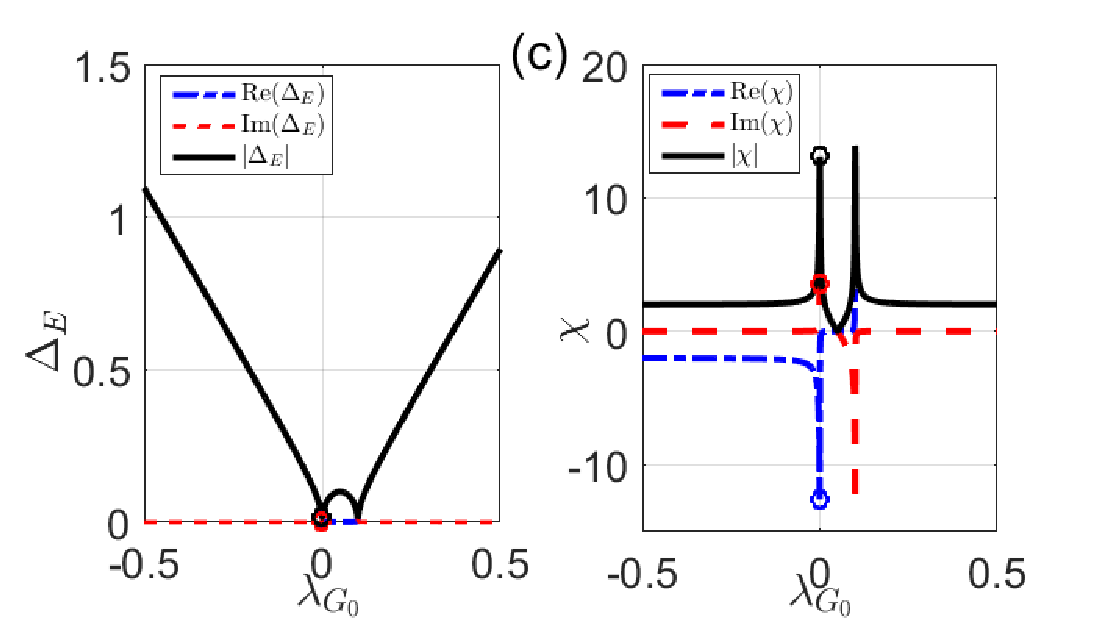}} \caption{Same
as Fig.~\ref{fig:E}, but evaluated at the fixed scaled time
$\theta/\pi=1$.} \label{fig:E-2D}
\end{figure}

Having established how parameter trajectories control the symmetry
and robustness of state transfer, we now turn to their
implications for quantum sensing. We first consider
eigenvalue-based sensing, which detects small parameter
perturbations via shifts in the system energy splitting. We
introduce the perturbation strength for a control parameter $X$ as
\begin{equation}\label{eq:X}
\lambda_X = X^{\text{R}} - X^{\text{I}}
\end{equation}
where $X^{\text{I}}$ and $X^{\text{R}}$ denote the ideal and
actual values, respectively. From Eq.~(\ref{eq:eigenvalues}), we
derive the susceptibilities as the derivatives of
$\Delta_E(\theta)$ with respect to $G_0$ and $\Gamma_0$:
\begin{subequations}\label{eq:susceptibilities}
\begin{align}
\chi_{G_0}(\theta) &=\frac{\partial\Delta_E}{\partial
g}\frac{\partial g}{\partial G_0}
= \frac{4g(\theta)}{\Delta_E(\theta)}\cos\theta, \\
\chi_{\Gamma_0}(\theta) &=
\frac{\partial\Delta_E}{\partial\gamma}\frac{\partial\gamma}{\partial\Gamma_0}
=
-\frac{\gamma(\theta)+i\delta(\theta)}{\Delta_E(\theta)}\sin^2(\theta/2).
\end{align}
\end{subequations}
Both susceptibilities diverge as $\Delta_E(\theta)\to0$, i.e.,
near an exceptional point
(EP)~\cite{Enhanced-sensitivity,sensing-microcavity}, which is the
origin of the ultrahigh sensitivity in EP-enhanced sensing.

Figure~\ref{fig:E} plots $\Delta_E(\theta)$ and $\chi(\theta)$
versus scaled time $\theta/\pi$ for the three trajectories under
weak perturbations. Sharp variations and near-zero crossings of
$\Delta_E(\theta)$ coincide with strong susceptibility peaks,
confirming the singular response near
EPs~\cite{Microcavity-sensor}. The 2D slices at $\theta/\pi=1$
(Fig.~\ref{fig:E-2D}) further show that the amplitude, width, and
position of $\chi(\theta)$ are strongly trajectory-dependent,
illustrating the tunability enabled by trajectory engineering.

We now analyze the three trajectories in terms of their
suitability for eigenvalue-based quantum sensing. We focus on
their sensitivity, selectivity, and practical utility. Trajectory
1 ($\delta=0$, with $g$ and $\gamma$ modulated): The absence of
detuning ($\delta=0$) eliminates complex phase effects that could
complicate the sensor response. The susceptibility $\chi_{G_0}$ is
proportional to $\cos\theta$, peaking at $\theta=0$ and
$\theta=\pi$. $\chi_{\Gamma_0}$ is scaled by $\sin^2(\theta/2)$,
peaking at $\theta=\pi$. Importantly, this trajectory does not
encircle the EP. Thus, the susceptibilities remain finite
(non-singular) but exhibit strong time dependence. This design
offers continuously tunable sensitivity with a smooth response
profile, making it well suited for applications requiring high
robustness against parameter fluctuations.

Trajectory 2 (constant $g$, $\gamma$ modulated, and finite
$\delta$): For this trajectory, since \(g(\theta) = 0.1\) is
constant and independent of \(G_0\) (i.e., \(\partial g / \partial
G_0 = 0\)), Eq. (12a) yields \(\chi_{G_0} = 0\). This means the
sensor is completely insensitive to fluctuations in \(G_0\). In
contrast, $\chi_{\Gamma_0}$ (evaluated at the ideal parameter
$\Gamma_0^{\text{I}} = 0.2$) is strongly enhanced by the complex
term $-(\gamma+i\delta)$. It exhibits sharp, singular peaks when
the parameter loop passes near the EP. This trajectory thus
realizes a single-parameter selective sensor. It responds
exclusively to variations in loss ($\Gamma_0$) while being immune
to noise in the coupling strength ($G_0$). This is a critical
feature for applications where selective detection is required
\cite{sensing-add1}.

Trajectory 3 (constant $\gamma$, $g$ modulated): Here,
$\chi_{\Gamma_0}=0$, rendering the sensor blind to fluctuations in
$\Gamma_0$. Conversely, $\chi_{G_0}$ (evaluated at the ideal
parameter $G_0^{\text{I}} = 0.15$) is strongly modulated by the
large-amplitude oscillations of $g(\theta)$. It yields broad
high-sensitivity windows near $\cos\theta\approx\pm1$. This design
supports wide-range, continuous detection of coupling
perturbations ($\lambda_{G_0}$) with relaxed timing constraints,
as the high-sensitivity regions span extended time intervals
\cite{quantum-sensing6}.

In summary, trajectory engineering enables precise control over
the directionality, selectivity, and temporal profile of the
eigenvalue-based sensor response. By tailoring the parameter
trajectory, one can design quantum sensors optimized for specific
detection tasks. These range from robust, general-purpose sensing
to highly selective, single-parameter detection.

\subsection{Eigenstate-based sensing}
\label{subsec:eigenstateSensing}

While eigenvalue-based sensing relies on spectral shifts, a
complementary and often more powerful approach is eigenstate-based
sensing, which monitors the population dynamics of targeted
eigenstates. We define the normalized population of the eigenstate
\(|\phi_{+}(t,\lambda)\rangle\) as
\begin{equation}\label{eq:population}
    P_{+}(t,\lambda)=\frac{|\langle\tilde{\phi}_{+}(t)|\Psi(t,\lambda)\rangle|^{2}}{|\langle\tilde{\phi}_{+}(t)|\Psi(t,\lambda)\rangle|^{2} + |\langle\tilde{\phi}_{-}(t)|\Psi(t,\lambda)\rangle|^{2}},
\end{equation}
where \(|\Psi(t,\lambda)\rangle\) denotes the time-evolved state
obtained by solving the time-dependent Schr\"{o}dinger's equation
for the perturbed parameter trajectory (\(\lambda\neq0\)). The
corresponding susceptibility, which quantifies the sensitivity of
the normalized population to small parameter perturbations
\(\lambda\), is defined as
\begin{equation}\label{eq:susceptibilityPop}
\tilde{\chi}(t,\lambda)=\frac{\partial
P_{+}(t,\lambda)}{\partial\lambda}
\approx\frac{P_{+}(t,\lambda_{i+1})-P_{+}(t,\lambda_{i})}{\lambda_{i+1}-\lambda_{i}}.
\end{equation}
Here, the finite-difference approximation is employed for
numerical practicality, as it avoids the need for analytical
differentiation of the population \(P_{+}(t,\lambda)\), which can
be computationally cumbersome for complex parameter trajectories.

\begin{figure}[htb]
\centering \scalebox{0.45}{\includegraphics{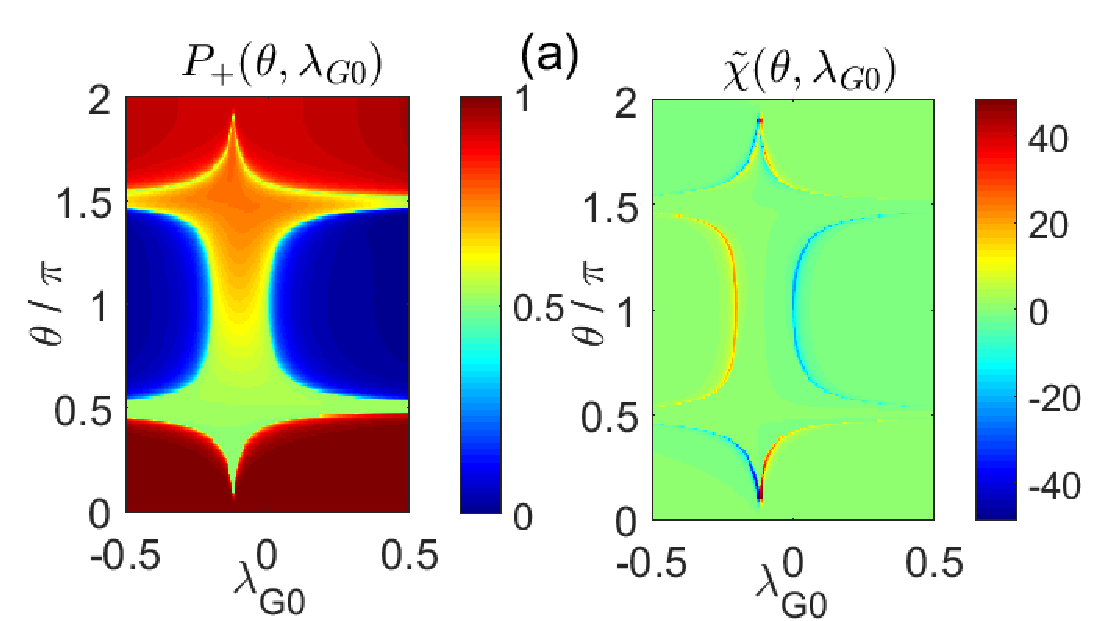}}
\scalebox{0.45}{\includegraphics{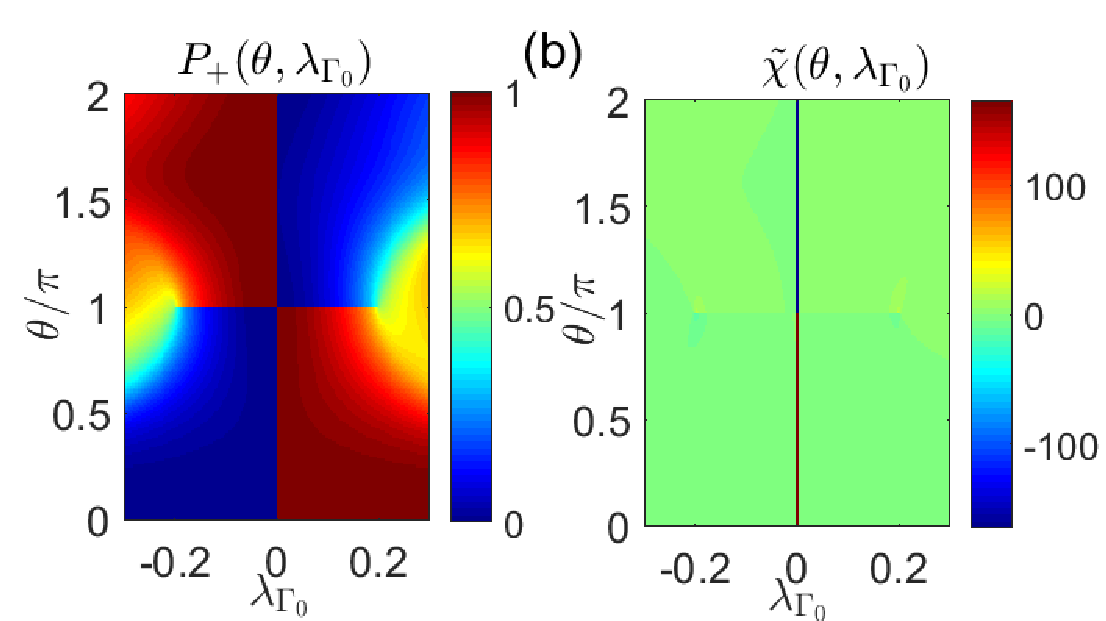}}
\scalebox{0.45}{\includegraphics{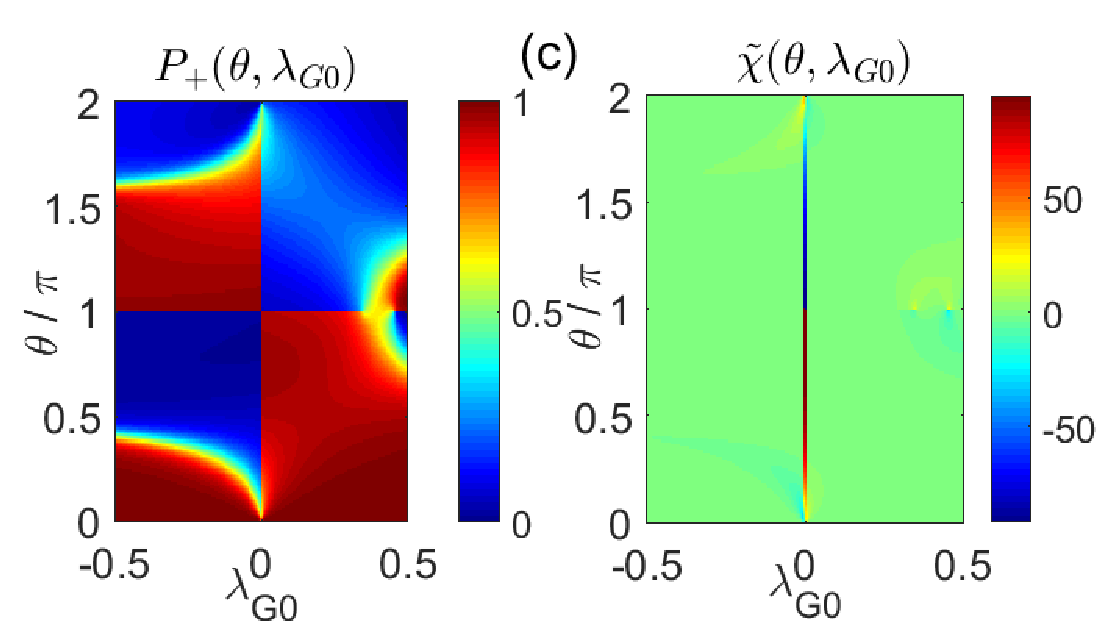}}
\caption{Normalized population $P_{+}(\theta,\lambda)$ and
susceptibility $\tilde{\chi}(\theta,\lambda)$ versus scaled time
$\theta/\pi$ for small control-parameter perturbations. (a)
Trajectory 1 with $G_0$ perturbation ($G_0^{\text{I}} = 0.11$);
(b) Trajectory 2 with $\Gamma_0$ perturbation
($\Gamma_0^{\text{I}} = 0$); (c) Trajectory 3 with $G_0$
perturbation ($G_0^{\text{I}} = -0.2$). The initial state is
$|\phi_{-}(0)\rangle$, and other parameters are given in
Eqs.~(\ref{eq:traj1})-(\ref{eq:traj3}).} \label{fig:P-3D}
\end{figure}

\begin{figure}[htb]
\centering \scalebox{0.45}{\includegraphics{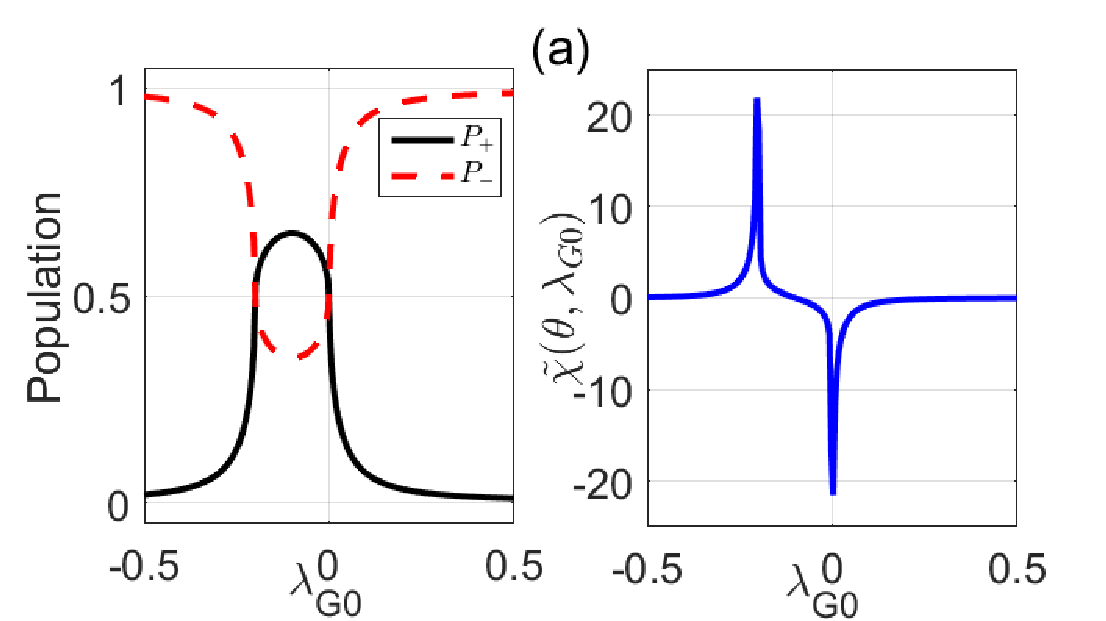}}
\scalebox{0.45}{\includegraphics{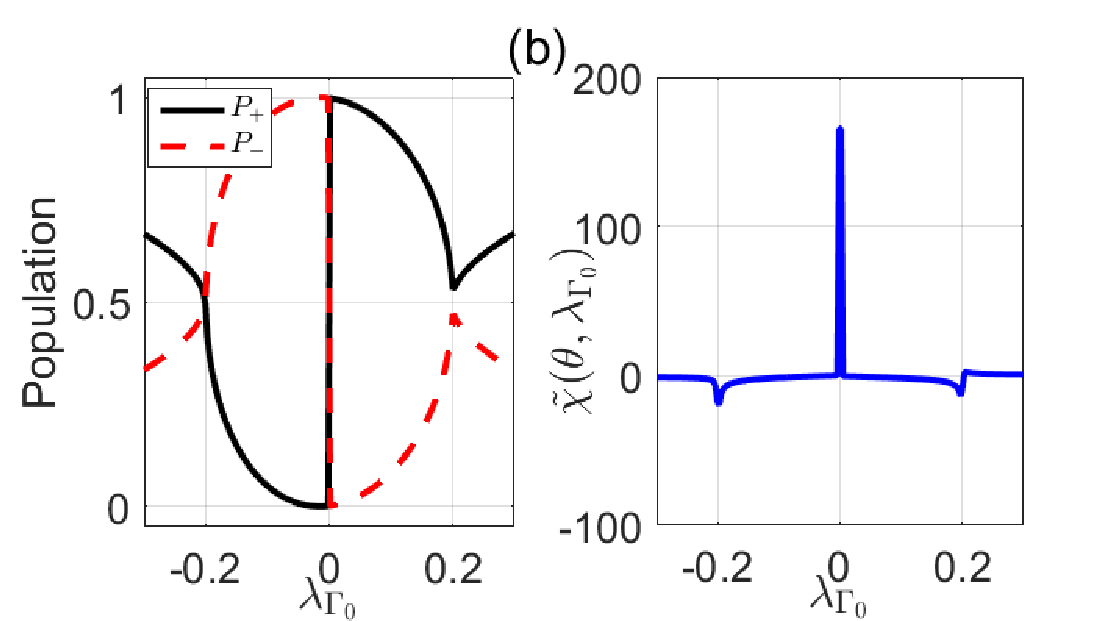}}
\scalebox{0.45}{\includegraphics{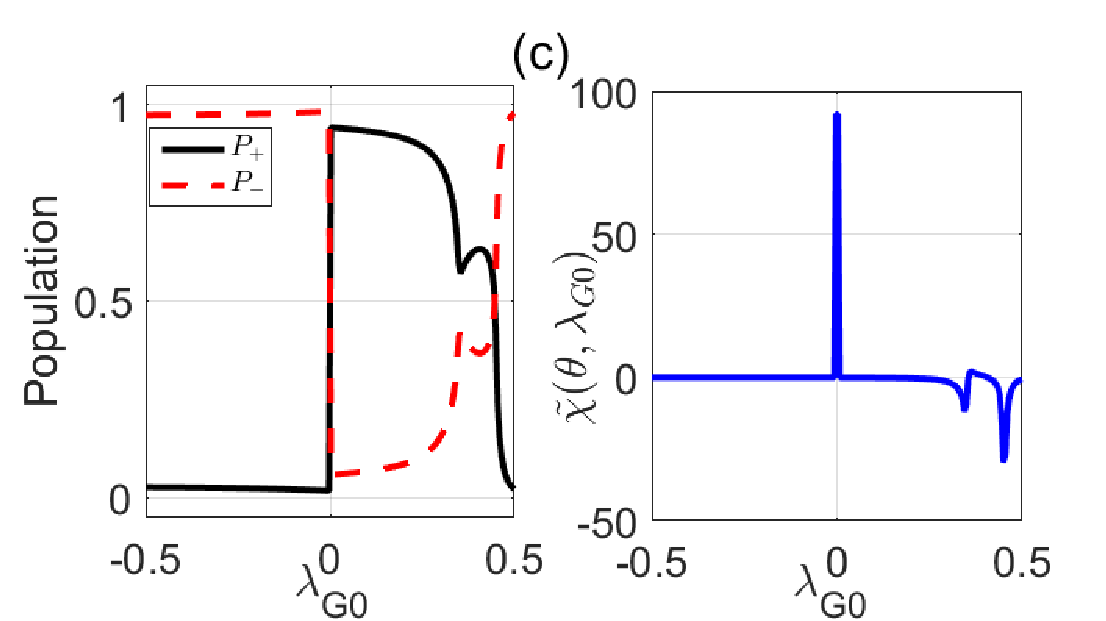}} \caption{Same
as Fig.~\ref{fig:P-3D}, but at the fixed scaled time
$\theta/\pi=1$.} \label{fig:P-2D}
\end{figure}

Figures~\ref{fig:P-3D} and~\ref{fig:P-2D} present the normalized
population \(P_{+}(\theta,\lambda)\) and the susceptibility
\(\tilde{\chi}(\theta,\lambda)\) for the three parameter
trajectories. The initial state is fixed as
\(|\Psi(0)\rangle=|\phi_{-}(0)\rangle\). The two-dimensional (2D)
slices at \(\theta/\pi=1\), corresponding to the completion of one
full parameter cycle, highlight the optimal instants for sensing
\cite{quantum-sensing4}. These time points capture the maximum
response of the population to parameter perturbations.

Trajectory 1: The normalized population \(P_{+}\) undergoes a
rapid transition near \(\theta/\pi\approx1\) and
\(\lambda_{G_0}\approx0\). This gives rise to a sharp
susceptibility peak with \(|\tilde{\chi}|\approx 20\). Consistent
with its non-chiral, topologically trivial nature (as established
in the preceding section), this trajectory enables precise sensing
of \(G_0\) fluctuations with high robustness. A key limitation,
however, is its narrow effective time window, which is centered
around the completion of the full parameter cycle.

Trajectory 2: The population $P_{+}$ exhibits quadrant-dependent
chiral evolution, which arises directly from the trajectory
encircling the EP \cite{topology-encircling}. The susceptibility
reaches a peak value $|\tilde{\chi}|\approx 160$ at $\theta/\pi=1$
and $\lambda_{\Gamma_0}\approx0$, evaluated at the ideal parameter
$\Gamma_0^{\text{I}} = 0$. Notably, if we set the ideal value for
$\Gamma_0$ in Trajectory 2 to $|\Gamma_0^{\text{I}}| \approx 0.2$,
the peak value of $|\tilde{\chi}|$ drops significantly to
approximately 20. This demonstrates that the modulation of
trajectory parameters strongly affects the sensitivity to a
specific parameter, even for the same parameter trajectory. On the
other hand, $\tilde{\chi}_{G_0}=0$ for this trajectory, making the
sensor highly selective to $\Gamma_0$ while being immune to
cross-talk from variations in the coupling strength $G_0$. This is
a key feature for applications requiring single-parameter
detection \cite{sensing-add1}.

Trajectory 3: The population $P_{+}$ exhibits a broad
high-sensitivity band. For instance, $|\tilde{\chi}|\approx 90$
near $\lambda_{G_0}=0$, $|\tilde{\chi}|\approx 15$ near
$\lambda_{G_0}=0.35$, and $|\tilde{\chi}|\approx 25$ near
$\lambda_{G_0}=-0.35$. Moreover, Trajectory 3 exhibits
$\tilde{\chi}_{\Gamma_0}=0$ (with $\Gamma=0.1$ held constant),
ensuring full immunity to fluctuations in the loss parameter
$\Gamma_0$. This design therefore achieves an excellent balance
among three key performance metrics: high sensitivity, a broad
operating time window, and strong anti-interference against
unintended loss perturbations.

The three trajectories exhibit distinct performance trade-offs for
eigenstate-based sensing. Trajectory 1 delivers moderate
sensitivity and strong robustness, yet is constrained by a narrow
effective time window. Trajectory 2 yields the highest peak
sensitivity and pronounced chiral response, but its sensitivity
depends sensitively on trajectory parameters and exhibits degraded
stability under noisy experimental conditions. In contrast,
Trajectory 3 achieves the most favorable overall performance by
maintaining high sensitivity over a broad temporal window,
together with strong immunity to unintended loss perturbations.

From these observations, we establish three key principles for
optimal trajectory design in eigenstate-based sensing: (i) the
trajectory should approach the EP to induce divergent
susceptibility for ultrahigh
sensitivity~\cite{Enhanced-sensitivity}; (ii) the trajectory
should enable independent control of $G_0$ and $\Gamma_0$ to
achieve full parameter selectivity~\cite{sensing-add1}; (iii) the
trajectory should sustain wide high-sensitivity temporal windows
to relax experimental timing requirements~\cite{quantum-sensing6}.
Among the three trajectories examined, Trajectory 3 satisfies all
three criteria and thus represents a better choice for practical
implementations.

Our results confirm that parameter-trajectory engineering provides
a unified framework for tailoring the susceptibility in
eigenstate-based quantum sensing. By combining EP-enhanced
sensitivity, chiral dynamical response, parameter selectivity, and
controllable sensing windows, this scheme offers a
high-performance platform for the precision detection of weak
physical signals in realistic quantum sensing
applications~\cite{quantum-sensing2,quantum-sensing3}.

\section{Discussion and Conclusion}
\label{sec:conclusion}

We have systematically investigated the role of parameter
trajectory engineering in controlling state-transfer symmetry and
quantum-sensing performance in a NH two-level system, with a focus
on EP encirclement and associated topology. Our key findings can
be summarized as follows.

First, parameter trajectory design directly governs the symmetry
and robustness of state transfer. Trajectories that do not
encircle the EP support symmetric (nonchiral) state transfer
characterized by the topological winding number \(\nu(\Gamma)=0\),
which is highly robust against parameter fluctuations and
insensitive to the encircling direction. In contrast, trajectories
that encircle the EP produce chiral state transfer with
\(\nu(\Gamma)=\pm1/2\), where the sign is set by the encircling
direction. The robustness of chiral transfer depends on the
trajectory's proximity to the EP: sharp transitions in fidelity
and winding number occur for trajectories close to the EP, while
smoother transitions and improved stability appear for
trajectories at moderate distances. These results establish a
quantitative connection between the topological invariant
\(\nu(\Gamma)\) and the dynamical outcome of state transfer,
showing that topological protection preserves the symmetry of
transfer against smooth parameter perturbations as long as
\(\nu(\Gamma)\) is preserved.

Second, trajectory engineering provides a powerful degree of
freedom to tailor quantum sensing performance. The sensitivity of
both sensing schemes is highly sensitive to the ideal values of
trajectory parameters: a moderate change in the ideal parameter
can drastically reduce the peak susceptibility, revealing the
critical role of precise parameter modulation. Both
eigenvalue-based (spectral shift detection) and eigenstate-based
(population monitoring) sensing schemes exploit the divergent
susceptibility near EPs. However, their overall performance,
including sensitivity amplitude, operating time window, and
parameter selectivity, depends strongly on the chosen trajectory.
Eigenvalue-based sensing enables tunable sensitivity with either
smooth (non-EP-encircling trajectories) or sharp (EP-proximal
trajectories) response profiles. Eigenstate-based sensing realizes
complete parameter selectivity, and the three trajectories show
distinct performance trade-offs: Trajectory 1 offers moderate
sensitivity and high robustness with a narrow effective time
window; Trajectory 2 delivers the highest peak sensitivity and
strong chirality but suffers from severe parameter dependence and
poor stability in noisy environments; Trajectory 3 optimally
balances high sensitivity, broad temporal windows, and strong
immunity to loss perturbations. Among the three trajectories
examined, Trajectory 3 achieves the best overall balance of high
sensitivity, broad operating windows, and strong anti-interference
capability, satisfying three key principles for optimal
eigenstate-based sensing: proximity to the EP, independent control
of coupling and loss, and relaxed timing constraints.

Third, we clarify the key distinction between the two sensing
modalities. Eigenvalue-based sensing relies on spectroscopic
readout to detect energy level shifts~\cite{Microcavity-sensor},
whereas eigenstate-based sensing requires state tomography to
monitor changes in eigenstate population~\cite{quantum-sensing3}.
This distinction carries important practical implications for
experimental implementation: eigenstate-based sensing offers
superior selectivity for single-parameter detection in noisy
environments.

Our work advances the fundamental understanding of NH system
dynamics by establishing a direct quantitative link between
parameter-trajectory topology, state transfer, and sensing
performance. The practical design principles derived herein
provide a unified framework for developing robust NH
state-transfer protocols and high-performance quantum sensors,
with potential applications in quantum information processing
(e.g., chiral state switches) and precision metrology (e.g.,
ultrasensitive detection of weak
perturbations)~\cite{EP_shift,linewidth}. Future work could extend
this analysis to higher-dimensional NH
systems~\cite{physscr2025higher}, explore non-sinusoidal
modulation profiles, and validate the proposed trajectories in
experimental platforms including cavity quantum electrodynamics,
integrated photonic systems, and superconducting
circuits~\cite{experiment-realize3,Arkhipov2026}. Such extensions
would further strengthen the practical relevance of parameter
trajectory engineering for next-generation NH quantum
technologies.

\section*{ACKNOWLEDGEMENT}
This work was supported by the National Key Research and
Development Program of China (Grant No.~2024YFA1408900), the
National Natural Science Foundation of China (NSFC) (Grants
Nos.~12264040, 12374333, and 12364048),
 the Jiangxi Natural Science Foundation (Grant
Nos.~20232BCJ23022, 20252BAC240119, and 20252BAC220006),  the
Jiangxi Province Key Laboratory of Applied Optical Technology
(Grant No.~2024SSY03051), the Shangrao City Science and Technology
Plan Project under Grant No. 2025D010, and the Innovation Program
for Quantum Science and Technology under Grant No. 2021ZD0301705.

\section*{DATA AVAILABILITY}

The data that support the findings of this article are openly
available~\cite{dates}.

\end{CJK*}

\begin{thebibliography}{99}
\bibitem{EPs1} C. M. Bender and S. Boettcher, Real Spectra in Non-Hermitian Hamiltonians Having PT Symmetry, Phys. Rev. Lett. \textbf{80}, 5243 (1998).
\bibitem{EPs2} W. D. Heiss, Repulsion of resonance states and exceptional points, Phys. Rev. E \textbf{61}, 929 (2000).
\bibitem{EPs3} H. Cartarius, J. Main, and G. Wunner, Exceptional points in atomic spectra, Phys. Rev. Lett. \textbf{99}, 173003 (2007).
\bibitem{square-root1} M. V. Berry, Physics of nonhermitian degeneracies, Czech. J. Phys. \textbf{54}, 1039 (2004).
\bibitem{square-root2} W. D. Heiss, The physics of exceptional points, J. Phys. A \textbf{45}, 444016 (2012).

\bibitem{Enhanced-sensitivity} H. Hodaei, A. U. Hassan, S. Wittek, H. Garcia-Gracia, R. El-Ganainy, D. N. Christodoulides, and M. Khajavikhan, Enhanced sensitivity at higher-order exceptional points, Nature \textbf{548}, 187 (2017).
\bibitem{sensing-microcavity} W. J. Chen, S. \"{O}zdemir, G. Zhao, J. Wiersig, and L. Yang, Exceptional points enhance sensing in an optical microcavity, Nature \textbf{548}, 192 (2017).
\bibitem{Microcavity-sensor} J. Wiersig, Enhancing the sensitivity of frequency and energy splitting detection by using exceptional points: Application to microcavity sensors for single-particle detection, Phys. Rev. Lett. \textbf{112}, 203901 (2014).

\bibitem{topology1} E. J. Bergholtz, J. C. Budich, and F. K. Kunst, Exceptional topology of non-Hermitian systems, Rev. Mod. Phys. \textbf{93}, 015005 (2021).
\bibitem{topology-encircling} C. Guria, Q. Zhong, S. K. \"{O}zdemir, Y. S. S. Patil, R. El-Ganainy, and J. G. E. Harris, Resolving the topology of encircling multiple exceptional points, Nat. Commun. \textbf{15}, 1369 (2024).

\bibitem{NH1} L. Feng, R. El-Ganainy, and L. Ge, Non-Hermitian photonics based on parity-time symmetry, Nat. Photon. \textbf{11}, 752 (2017). \bibitem{NH2} Y. Ashida, Z. Gong, and M. Ueda, Non-Hermitian physics, Adv. Phys. \textbf{69}, 249 (2020).
\bibitem{NH3} Q. C. Wu, J. L. Zhao, Y. L. Fang, Y. Zhang, D. X. Chen, C. P. Yang, and F. Nori, Extension of Noether's theorem in PT-symmetry systems and its experimental demonstration in an optical setup, Sci. China-Phys. Mech. Astron. \textbf{66}, 240312 (2023).

\bibitem{wu-encircling} Q. C. Wu, J. L. Zhao, Y. H. Zhou, B. L. Ye, Y. L. Fang, Z. W. Zhou, and C. P. Yang, Shortcuts to adiabatic state transfer in time-modulated two-level non-Hermitian systems, Phys. Rev. A \textbf{111}, 022410 (2025).
\bibitem{Arkhipov-encircling2} I. I. Arkhipov, A. Miranowicz, F. Minganti, S. \"{O}zdemir, and F. Nori, Restoring adiabatic state transfer in time-modulated non-hermitian systems, Phys. Rev. Lett. \textbf{133}, 113802 (2024).

\bibitem{PhysRevLett.129.127401}Y. V. Kartashov, V. V. Konotop, and D. A. Zezyulin, Chiral State Transfer by Sequential Encircling of Exceptional Points, Phys. Rev. Lett. \textbf{129}, 127401 (2022).
\bibitem{PhysRevLett.124.153903} Q. Liu, B. Hou, Z. H. Jiang, X. Lin, and J. Wang, Efficient Mode Transfer on a Compact Silicon Chip by Encircling Moving Exceptional Points, Phys. Rev. Lett. \textbf{124}, 153903 (2020).
\bibitem{bell-encircling} S. Khandelwal, W. J. Chen, K. W. Murch, and G. Haack, Chiral Bell-State Transfer via Dissipative Liouvillian Dynamics, Phys. Rev. Lett. \textbf{133}, 070403 (2024).
\bibitem{Xu-encircling} H. Xu, D. Mason, Luyao Jiang, and J. G. E. Harris, Topological energy transfer in an optomechanical system with exceptional points, Nature (London) \textbf{537}, 80 (2016).
\bibitem{Li-encircling} A. Li, J. Dong, J. Wang, Z. Cheng, J. S. Ho, D. Zhang, et al., Hamiltonian hopping for efficient chiral mode switching in encircling exceptional points, Phys. Rev. Lett. \textbf{125}, 187403 (2020).
\bibitem{Feilhauer-encircling} J. Feilhauer, A. Schumer, J. Doppler, A. A. Mailybaev, J. B\"{o}hm, U. Kuhl, N. Moiseyev, and S. Rotter, Encircling exceptional points as a non-Hermitian extension of rapid adiabatic passage, Phys. Rev. A \textbf{102}, 040201 (2020).
\bibitem{Ergoktas-encircling}M. S. Ergoktas, S. Soleymani, N. Kakenov, K. Wang, T. B. Smith, G. Bakan, S. Balci, A. Principi, K. S. Novoselov, S. K. Ozdemir, and C. Kocabas, Topological engineering of terahertz light using electrically tunable exceptional point singularities, Science \textbf{376}, 184 (2022).
\bibitem{Arkhipov-encircling1} I. I. Arkhipov, A. Miranowicz, F. Minganti, S. \"{O}zdemir, and F. Nori, Dynamically crossing diabolic points while encircling exceptional curves: A programmable symmetric-asymmetric multimode switch, Nat. Commun. \textbf{14}, 2076 (2023).
\bibitem{Tang-encircling} Z. Tang, T. Chen, and X. Zhang, Highly efficient transfer of quantum state and robust generation of entanglement state around exceptional lines, Laser Photonics Rev. \textbf{17}, 2300794 (2023).
\bibitem{Hassan-encircling2} A. U. Hassan, B. Zhen, M. Solja\v{c}i\'{c}, M. Khajavikhan, and D. N. Christodoulides, Dynamically encircling exceptional points: Exact evolution and polarization state conversion, Phys. Rev. Lett. \textbf{118}, 093002 (2017).
\bibitem{negativecavity2} X. Tang, T. Chen, and X. D. Zhang, Controlling transfer and chirality of topological quantum state through dissipation in quantum walk, Phys. Rev. Research \textbf{7}, 013159 (2025).

\bibitem{PhysRevX.8.021066} X. L. Zhang, S. Wang, B. Hou, and C. T. Chan, Dynamically Encircling Exceptional Points: In Situ Control of Encircling Loops and the Role of the Starting Point, Phys. Rev. X \textbf{8}, 021066 (2018).

\bibitem{Arkhipov2026} I. I. Arkhipov, F. Nori, and S. K. \"{O}zdemir, Achieving the Quantum Fisher Information Bound in Pseudo-Hermitian Sensors, Phys. Rev. Lett. \textbf{136}, 080802 (2026).

\bibitem{quantum-sensing2} C. L. Degen, F. Reinhard, and P. Cappellaro, Quantum sensing, Rev. Mod. Phys. \textbf{89}, 035002 (2017).
\bibitem{quantum-sensing3} M. Z. Zhang, W. Sweeney, C. W. Hsu, L. Yang, A. D. Stone, and L. Jiang, Quantum Noise Theory of Exceptional Point Sensors, Phys. Rev. Lett. \textbf{123}, 180501 (2019).
\bibitem{quantum-sensing4} J. Wiersig, Prospects and fundamental limits in exceptional point-based sensing, Nat. Commun. \textbf{11}, 2454 (2020).
\bibitem{quantum-sensing5} Q. C. Wu,  Y. H. Zhou,  T. Liu,  Y. H. Kang,  Q. P. Su, and C. P. Yang,  Enhanced quantum sensing in time-modulated non-Hermitian systems, Chin. J. Phys. \textbf{98} 1116 (2025).
\bibitem{quantum-sensing6} W. C. Wong and J. S. Li, Exceptional-point sensing with a quantum interferometer, New J. Phys. \textbf{25}, 033018 (2023).
\bibitem{sensing-add1} J. N. Li, H. D. Liu, Z. H. Wang, and X. X. Yi, Enhanced parameter estimation by measurement of non-Hermitian operators, AAPPS Bulletin \textbf{33}, 22 (2023).
\bibitem{experiment-realize3} H. S. Xu and L. Jin, Coupling-induced nonunitary and unitary scattering in anti-PT-symmetric non-Hermitian systems, Phys. Rev.A \textbf{104}, 012218 (2021).

\bibitem{PhysRevLett.127.253901} J. Li, A. K. Harter, J. Liu, L. de Melo, Y. N. Joglekar, and L. Luo, Non-Abelian Geometric Phase from Dynamical Encirclement of Exceptional Points, Phys. Rev. Lett. \textbf{127}, 253901 (2021).

\bibitem{JC2}Y. H. Zhou, H. Z. Shen, X. Y. Zhang, and X. X. Yi, Zero eigenvalues of a photon blockade induced by a non-Hermitian Hamiltonian with a gain cavity, Phys. Rev. A \textbf{97}, 043819 (2018).
\bibitem{wu-encircling1}Q. C. Wu, Y. L. Fang, Y. H. Zhou, J. L. Zhao, Y. H. Kang, Q. P. Su,  C. P. Yang, Efficient and controlled symmetric and asymmetric Bell-state transfers in a dissipative Jaynes-Cummings model, Chin. Phys. B \textbf{35}, 010304 (2026).

\bibitem{PhysRevLett.129.273601} Y. V. Kartashov, D. A. Zezyulin, and V. V. Konotop, Topological Protection and Switchable Chirality in Non-Hermitian Systems, Phys. Rev. Lett. \textbf{129}, 273601 (2022).
\bibitem{PhysRevLett.132.243802} J. Doppler, S. Rotter, and A. A. Mailybaev, Non-Abelian Topological Transitions via Exceptional Point Encirclement, Phys. Rev. Lett. \textbf{132}, 243802 (2024).

\bibitem{tp1} J. W. Ryu, J. H. Han, and C. H. Yi, Realization of geometric-phase topology induced by multiple exceptional points, Phys. Rev. A \textbf{110}, 052221 (2024).
\bibitem{tp2} D. Leykam, K. Y. Bliokh, C. Huang, Y. D. Chong, and F. Nori, Edge modes, degeneracies, and topological numbers in non-hermitian systems, Phys. Rev. Lett. \textbf{118}, 040401 (2017).

\bibitem{EP_shift} X. Mao, G. Q. Qin, H. Zhang, B. Y. Wang, D. Long, G. Q. Li, and G. L. Long, Enhanced sensing mechanism based on shifting an exceptional point, Research \textbf{6}, 0260 (2023).
\bibitem{linewidth} D. Long, J. D. Zhu, X. Mao, G. Q. Qin, M. Wang, G. Q. Li, F. Bo, and G. L. Long, Exceptional-point-enhanced nanoparticle sensor utilizing a linewidth broadening mechanism, Opt. Lett. \textbf{50}, 852 (2025).

\bibitem{physscr2025higher} A. Roy, A. Laha, A. Biswas, B. P. Pal, S. Ghosh, and A. Miranowicz, Dynamically encircled higher-order exceptional points in an optical fiber, Phys. Scr. \textbf{100}, 045529 (2025).

\bibitem{dates} Q. C. Wu, Y. H. Zhou, B. L. Ye, T. Liu, Y. H. Kang, Q. P. Su,  C. P. Yang, code for this
study, https://github.com/wuqicheng-ui/ep-state-transfer.

\end{thebibliography}
\end{document}